\documentclass[reprint, prb, amsmath, amssymb, nofootinbib]{revtex4-2} 			 
\usepackage{graphicx}
\usepackage{pifont}
\usepackage{bm} 
\usepackage{dcolumn} 
\usepackage[colorlinks=true, citecolor=blue, linkcolor=blue]{hyperref}%
\usepackage[normalem]{ulem}  

\allowdisplaybreaks

\begin{document}
\title{Magnetic field study of exciton nonradiative broadening excitation spectra in GaAs/AlGaAs quantum wells}


\author{M.\,A. Chukeev}
\author{A.\,S. Kurdyubov}
\author{I.\,I.  Ryzhov}
\author{P.\,S. Grigoryev}
\affiliation{Spin Optics Laboratory, St. Petersburg State University, Ulyanovskaya 1, Peterhof, St. Petersburg, 198504, Russia}

\author{V.\,A. Lovtcius}
\author{Yu.\,P. Efimov}
\author{S.\,A. Eliseev}
\affiliation{Research Center "Nanophotonics", St. Petersburg State University, Ulyanovskaya 1, Peterhof, St. Petersburg, 198504, Russia}

\begin{abstract}
Exciton excited states in the quantum well are studied via their effect on the nonradiative broadening of the ground exciton resonance in the reflectance spectrum. Dependence of the nonradiative broadening of the ground exciton state on the photon energy of additional laser excitation was measured. Applying magnetic field up to 6\,T, we could trace the formation of Landau levels and evolution of the exciton states of size quantization in a 14-nm GaAs/AlGaAs quantum well. Sensitivity of the technique allowed for observation of the second exciton state of size quantization, unavailable for conventional reflectance and photoluminescence spectroscopy. Our interpretation is supported by the numerical calculation of the exciton energies of the heavy-hole and light-hole subsystems. The numerical problems were solved using the finite-difference method on the nonuniform grid. The ground Landau level of the free electron-hole pair was observed and numerically analysed. In addition to energies of the excited states, electron hole distances and exciton-light interaction constant was investigated using the obtained in the numerical procedure exciton wave functions.
\end{abstract}
\maketitle 

\section{Introduction}

The persistent advancements in molecular-beam epitaxy (MBE) have enabled the growth of high-quality GaAs-based nanostructures with quantum wells (QW). Reflectance spectroscopy of these QWs reveal exciton resonances of comparable radiative and nonradiative broadening values~\cite{Trifonov-PRB2015}. This has paved the way for the development of a novel technique called the nonradiative broadening excitation spectroscopy for study of exciton scattering (NBE spectroscopy)~\cite{Kurdyubov-PRB2021}, which is based on the measurement of the nonradiative broadening of the exciton resonance under an additional excitation by a tunable laser.

In Ref.~\cite{Kurdyubov-PRB2021} the NBE technique has been used to investigate the subbands of free charges carriers in the high-quality nanostructures. The results show that these subbands manifest themselves as step-like features in the NBE spectra. This work is aimed to investigate the evolution of the NBE spectra under external magnetic fields. Landau levels are expected to form from the carriers' subbands. The investigation of the NBE spectra under external magnetic fields will provide a better understanding of the physics behind the formation of Landau levels and the behavior of excitons subjected to the magnetic field.

Exciton energy shift in magnetic field was a decisive argument proving it's nature of the bonded electron-hole pair at the time of it's experimental discovery~\cite{Gross}. With QWs emerged~\cite{Dingle-PRL1974, Deveaud-PRL1991, Andreani-SSC1991}, the excited hydrogen-like exciton states would only be observable in the so-called thin layers~\cite{Kusano-SSC1989}, QWs with width much larger than the exciton Bohr radius. However, for the wider QWs the magnetooptical studies would be focused on intriguing $g$-factor renormalisation effect for the states of size quantisation~\cite{Davies-PRB2006, Smith-PRB2008, Davies-PRB2010, Smith-PRB2011}. This is related to the fact, that states of size quantisation are the dominating features in the optical reflectance or absorption spectra. 

In bulk materials modern experimental capabilities allow for meticulous study of excited hydrogen-like exciton states~\cite{Kazimierczuk-Nature2014, Thewes-PRL2015}. In this work we were able to study both exciton states of size quantisation and excited hydrogen-like exciton states. Studied QW has width $L_{QW}=14$\,nm, which is close to the exciton Bohr radius in GaAs. So the size quantisation effect is expected to be comparable to the Coulomb coupling effects, and we rely on the numerical calculation to distinguish the behavior of various exciton states under the magnetic field. For such conventional QWs, the effect of magnetic field taking into account the complex valence band nature was addressed multiple times~\cite{Durnev-PhysE2012, Arora-JAP2013, Timofeev-JETPL1996, Chen-APL2006}. However, to the best of our knowledge, excited exciton states have never been considered in detail.

In this paper, we present the results of our investigation of the NBE spectra under external magnetic field in Faraday geometry. Our results show that the NBE spectra exhibit distinct features, which we attribute to particular exciton states and Landau levels formed by free charge carriers. To achieve this we analyzed the numerically obtained exciton wave functions. A finite-difference method was used to convert a multidimensional Schroedinger equation into an eigenvalue problem for the large sparse matrix \cite{Grigoryev-PRB2016, Grigoryev-PRB2017, Khramtsov-JAP2016, Grigoryev-SuperMicro2016, Khramtsov-PRB2019, Belov-PhysE2019}. An underlying Hamiltonian accounts for the finite QW potential, the electron-hole Coloumb attraction, and heavy-hole-light-hole (hh-lh) coupling using the Luttinger Hamiltonian. Magnetic field effects are also covered by this approach.

The manuscript is arranged as follows: we first present the experimental and theoretical methods used in our study, followed by an analysis of our numerical results in comparison to experimental data, concluding with a discussion section and conclusions.

\section{Methods}

The sample under study contains a 14-nm GaAs QW, sandwiched between 50-nm AlGaAs barriers with 3\,\% Al content. Sample has a silicon doped GaAs substrate, which is separated from the QW and barriers by the 600-nm MBE-grown GaAs layer. On top of these layers, a 175\,nm 3\,\% AlGaAs layer was grown surrounded by 2.8-nm AlAs layers. A capping GaAs layer of 7\,nm width was grown on top of the sample. The layers on top of the QW serve two purposes. First, under normal incidence they provide the $\pi$ phase shift of the light reflected from the QW relative to the light reflected from the sample surface. Such phase shift results in a peak-shaped resonance form in the reflectance spectrum as we show below. Second, the capping layer is made mainly of the barrier material with a wider band gap, thus we eliminate the absorption of light on the way to the QW up to the barriers band gap photon energy.

In this paper we employ a recently developed method~\cite{Kurdyubov-PRB2021} of nonradiative broadening spectroscopy, for the sample under influence of the magnetic field. The reflectance spectrum of the sample containing a GaAs/AlGaAs QW is measured using conventional technique. Light of the incandescent lamp is filtered through a 100-$\mu$m pin hole and focused onto the sample placed in a closed-cycle cryostat with a superconducting split-coil magnet. The sample temperature is 1.5\,K. The reflected beam is collected in two circular polarizations in a 0.5-meter monochromator with 1200 gr./mm grating, and a CCD. The reflectance spectrum is measured while the wavelength of an additional continuous wave (CW) laser is scanned over the spectral range studied. The laser beam is circularly polarized and focused on the same point on the sample as the incandescent lamp radiation. The spectrum is detected in the cross-circular polarisation with respect to the laser beam to filter the stray laser radiation from the reflectance spectrum detected. At each laser wavelength, three spectra were measured: the reflectance spectrum, the reflectance spectrum with the additional excitation and the photoluminescence (PL) spectrum induced by the additional excitation. An automatic shutter system was built at the setup to open and close the lamp and laser beams. These measurements were conducted for magnetic field values up to 6\,T applied in the Faraday geometry (along the growth axis of the sample) with the 1\,T step.

\begin{figure}
    \centering
    \includegraphics[width=8.6cm]{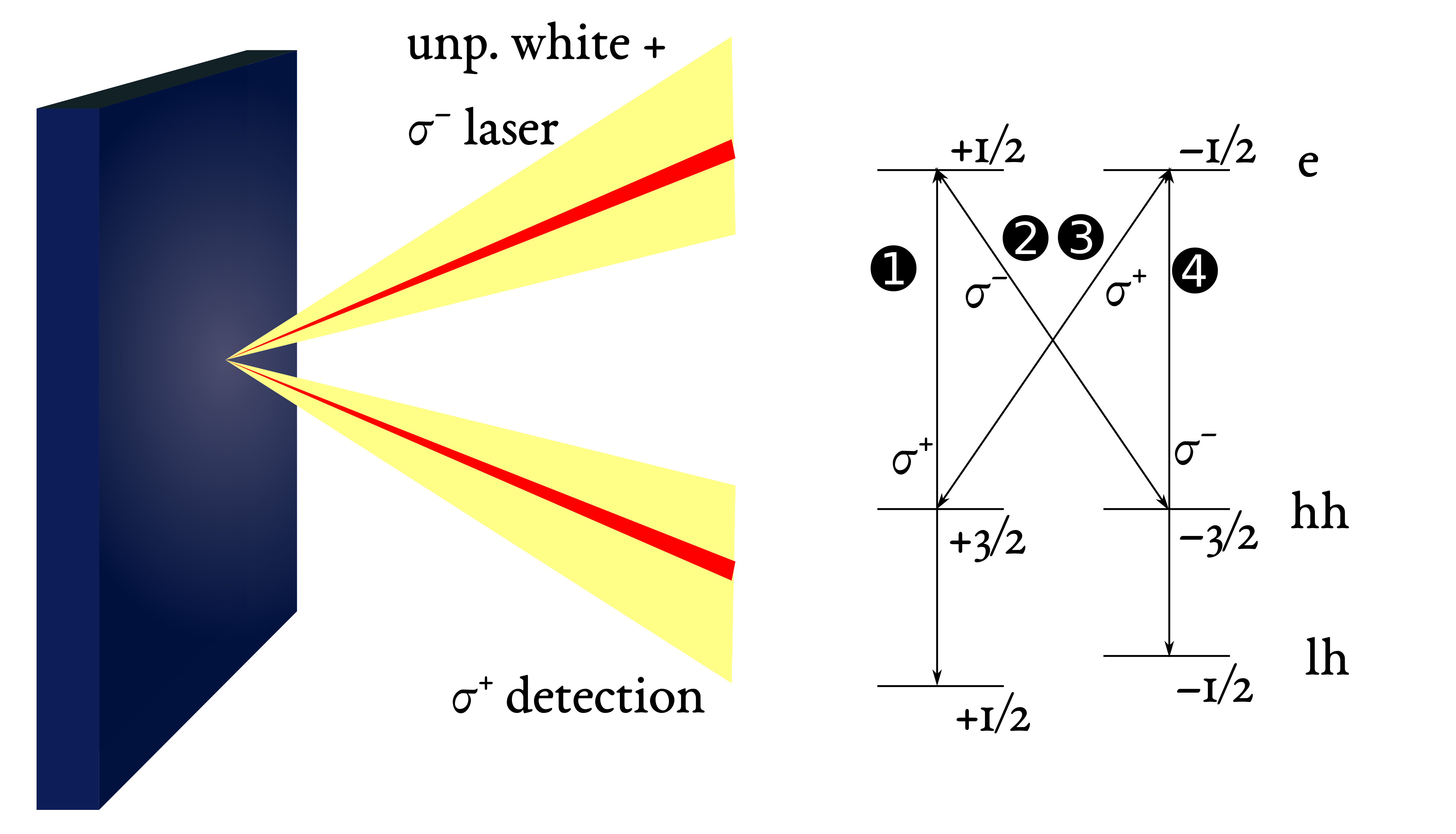}
    \caption{Excitation scheme used in our experiment. beams of white light and circularly polarized tunable laser fall on the sample. The reflected beams in opposite circular polarisation are detected. }
    \label{fig:scheme}
\end{figure}

The experimental data obtained was processed as follows. A PL spectrum was subtracted from the reflectance spectrum with additional laser excitation, and the exciton resonances were fitted according to the formula~\cite{Ivchenko-book2004}:

\begin{equation}
    R=\left|\frac{r_0+e^{i \phi} r_\text{QW}}{1+r_0 e^{i \phi} r_\text{QW}}\right|^2 \label{refl}
\end{equation}
with $r_0$ standing for the amplitude reflectance coefficient from the sample surface, and $\phi$ being the phase delay acquired by the light propagating to the QW and back to the sample surface. $r_\text{QW}$ stands for the amplitude reflectance coefficient of the QW:

\begin{equation}
    r_\text{QW}=\frac{i \Gamma_0}{\omega-\omega_0-i(\Gamma_0+\Gamma)}.
\end{equation}
Here $\omega_0$ is the energy of the exciton resonance, $\Gamma_0$ is the exciton radiative decay rate, and $\Gamma$ is the nonradiative decay rate. The latter two values define the broadening of the exciton resonance in reflectance spectrum.  The radiative component of the broadening, $\hbar \Gamma_0$, originates from the exciton radiative lifetime. The nonradiative broadening, $\hbar \Gamma$,  describes the scattering of the excitons on each other, impurities, phonons, free carriers, etc. In our analysis of the experimental data for a single sweep of the laser wavelength, the radiative broadening is considered unchanged because of the weak excitation~\cite{Mursalimov-FTP2022}, while the nonradiative broadening is a fitting parameter. Dependence of the nonradiative broadening of the exciton resonance on the energy of additional excitation we call a spectrum of the nonradiative broadening excitation (NBE)~\cite{Kurdyubov-PRB2021}.

\begin{figure}
    \centering
    \includegraphics{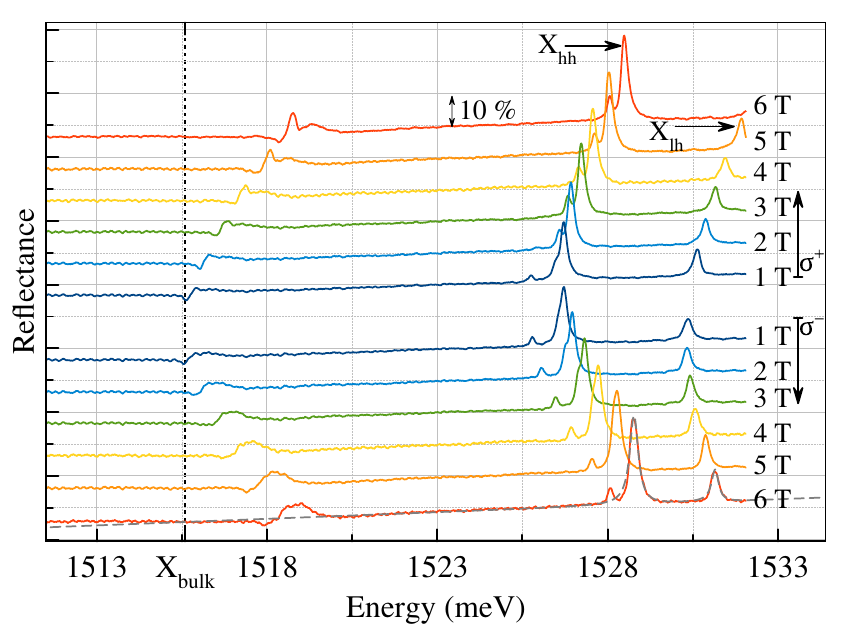}
    \caption{Reflectance spectra of the studied GaAs/AlGaAs heterostructure measured at magnetic field up to 6\,T in opposite circular polarisations. The gray dashed line along the bottom curve presents a typical fit of one of the spectrum with formula~(\ref{refl})}.
    \label{fig:refl}
\end{figure}

Figure~\ref{fig:refl} presents the reflectance spectra measured in our experiment. The exciton resonance in the buffer GaAs layer, $X_\text{bulk}$, is observed at $E_\text{bulk}=1515.6$\,meV without magnetic field. The heavy-hole and light-hole exciton resonances are denoted as "$X_\text{hh}$" and "$X_\text{lh}$". Below the $X_\text{hh}$ resonances a small peculiarity is observed that we tend to ascribe to a trion state. This "trion" is visible in $\sigma^-$ polarisation, and vanishes in opposite polarisation with increasing magnetic field. This behavior indicates that trion g-factor is negative and the hole relaxes to the $-3/2$ state in the magnetic field. In the left-hand polarisation, an additional resonance splits off from the heavy-hole resonance. It possibly can be dark exciton state, which acquires oscillator strength due to hh-lh coupling.

The selection rules for the circular polarisation are depicted in figure~\ref{fig:scheme}. Lets apply them for the case when the laser excites a lh exciton resonance in the left circular polarisation that is the transition denoted as \ding{205} in Fig.~\ref{fig:scheme}. The excitation populates the nonradiative reservoir with unpolarised excitons, because their polarisation is lost in a fraction of nanosecond~\cite{Trifonov-FTP2019}. These excitons affect the broadening of all the resonance measured in reflectance spectra in both the $\sigma^+$ and $\sigma^-$ polarisations. In what follows, we denote the polarisation of NBE spectra by the polarisation of the exciting laser light, that is by $\sigma^-$ in the considered case, rather than by the polarisation of the detected signal.

\begin{figure}
    \centering
    \includegraphics{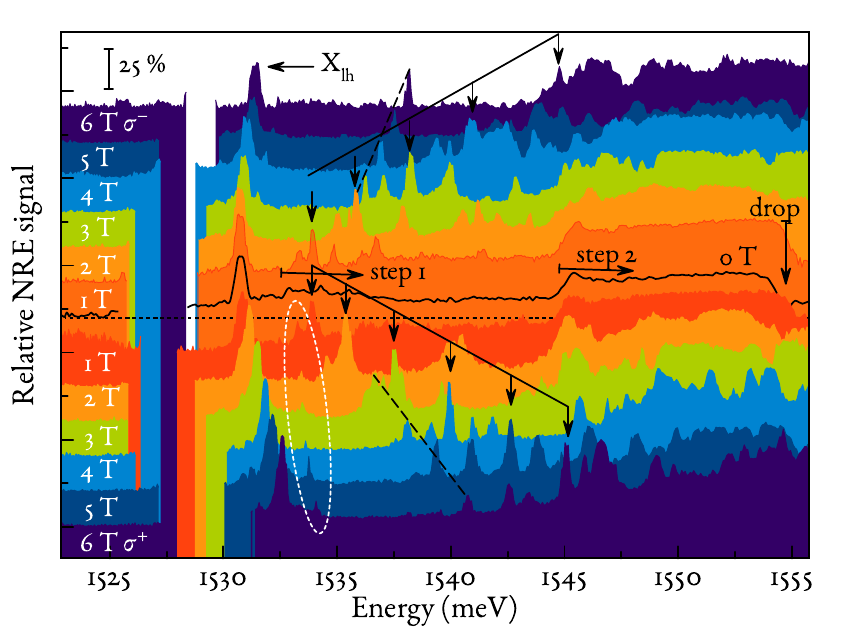}
    \caption{NBE spectra obtained in our experiments for $\sigma^+$ and $\sigma^-$ detection polarisation. The lh exciton resonance is denoted "X$_\text{lh}$". The black dashed lines and the solid black lines with arrows are guides for the eye to denote particular features in the spectra. The white oval selects feature, which we ascribe to the second hh exciton state of size quantuzation in the QW.}
    \label{fig:NBE_overview}
\end{figure}

As we show in the appendix, the NBE spectra can be measured in the broadening and in the narrowing regimes, depending on the measurement design. The excitation with a pulsed Ti:Sapphire laser provides initially narrow resonances in the reflectance spectra, and the excitation by an additional tunable laser increases the broadening. Measuring reflectance spectra with a white light source, provides a broader exciton resonances in reflectance, which are then exhibit narrowing under monochromatic excitation. As we found experimentally (see Appendix) same difference in the broadening is achieved at lower excitation laser power for the narrowing regime. We therefore chose this type of measurements due to limited power of the CW laser. 

Figure~\ref{fig:NBE_overview} presents relative nonradiative broadening change as a function of the photon energy of the CW excitation at various magnetic field. The absolute value of the difference in broadening imposed by the CW excitation is divided by the background nonradiative broadening in this figure. The black curve presents the NBE spectrum without the magnetic field. The region in the vicinity of the hh exciton resonance is unavailable due to the stray light of the laser, which is significantly suppressed by the polarisation selection, but can not be totally omitted due to the small aperture of the sample. The lh exciton state is present in spectrum as a dominant feature, which is followed by the two step-like features that correspond to the two subbands of free electron-hole pairs. These are marked as "step 1" and "step 2" in the figure. When magnetic field is applied, the spectra exhibit multiple distinct features that evolve in magnetic field. Further details in this figure and the particular description of observed features are discussed in section~\ref{sec:results}.

\subsection{Numerical modeling}
\label{sub:nummod}

As figure~\ref{fig:NBE_overview} shows, the NBE spectra measured in the magnetic fields demonstrate multiple features arising from the free electron-hole pairs being confined by the magnetic field. These are discrete states of Landau levels and exciton excited states, which can be hard to distinguish one from the other. This is because the Landau levels are still the subject to the Coulomb interaction. To sort out the states observed in NBE spectra, we employ two approaches to the numerical modeling. First one relies on the calculation of the energies and corresponding wave functions of the Landau levels of the electrons and holes. The second one calculates the excited exciton states exposed to the external magnetic field.

Both approaches solve numerically the Schroedinger equation. We used a finite difference method to produce a Hamiltonian matrix, which then was partially diagonalized using an Arnoldi algorithm. The two approaches differ in the used Hamiltonians. Both approaches however account for the hh-lh coupling.

To calculate the Landau levels we used a Hamiltonian of the form:

\begin{eqnarray}
    &\hat{H}_e= \frac{\hat{k}^2_e}{2 m_e}+U_c(z_e) \label{He},&\\
    &\hat{H}_h= \hat{H}_L(\hat{k}_h)+U_v(z_h) \label{Hh},&
\end{eqnarray}
with generalized momentum operators:
\begin{eqnarray}
&\hat{k}_e=-i\hbar\nabla_e+\frac{e}{2 c}\left[B\times r\right], \label{mome}&\\
&\hat{k}_h=-i\hbar\nabla_h-\frac{e}{2 c}\left[B\times r\right], \label{momh}&
\end{eqnarray}
where we use symmetric gauge. Second terms in equations~(\ref{He}) and~(\ref{Hh}) are the QW square potentials, and $\hat{H}_L$ is the Luttinger $4\times 4$ Hamiltonian. These Hamiltonians are three-dimensional and we use cylindrical coordinates to compose the Hamiltonian matrix on the 60\,nm $\times 400$\,nm domain along $z_{e,h}$ and $\rho_{e,h}$ coordinates with treating the angular coordinate via quantum number $m$. Detailed description of our approach to the Landau levels is discussed in section~\ref{sub:app_Ll}.

A second approach deals with numerical calculation of the Schroedinger equation with Hamiltonian taking into account the Coulomb interaction for electron-hole pair:

\begin{equation}
    \hat{H}_{ex}=\hat{H}_e+\hat{H}_h-\frac{e^2}{\varepsilon \sqrt{\rho^2+(z_e-z_h)^2}} \label{exham}
\end{equation}
here $\rho$ is the relative electron-hole distance in the plane perpendicular to the growth axis, and $\varepsilon$ is the dielectric constant. In the case of the exciton, however, we used the ansatz suggested by Gorjkov and Dzjaloshinskiy~\cite{Gorjkov-ZETP1967}. It allows for the center-of-mass coordinate separation in the Luttinger Hamiltonian with generalized momentum in form~(\ref{mome},\ref{momh}):

\begin{eqnarray}
    \psi=\exp{\left[i\frac{e B}{2 c \hbar}(Y x-X y)\right]}e^{i k_\phi \phi}\psi(z_e,z_h,\rho).
\end{eqnarray}

Here we again treat the part of the wave function depending on the angular relative coordinate $\phi$ in terms of quantum number $k_\phi$. This ansatz transforms Hamiltonian~(\ref{exham}) from $4\times 4$ matrix form into an infinite matrix form, corresponding to the infinite values of the $k_\phi$ number. Coupling between the heavy-hole (hh) state and the light-hole (lh) states in the Hamiltonian obeys the selection rules, which we explored previously~\cite{Grigoryev-PRB2016}. We found that the hh states couple with lh states with same angular momentum direction if their $k_\phi$ number differs by 1, and with opposite angular momentum projection if their $k_\phi$ number differs by $\pm 2$. Based on these rules we restrict the basis for the $\phi$ coordinate to 4 states: $\left|\text{hh},k_\phi=0\right>$,$\left|\text{lh},k_\phi=1\right>$,$\left|\text{lh},k_\phi=2\right>$, and $\left|\text{lh},k_\phi=-2\right>$. Separately we solved similar problem for the lh exciton with states $\left|\text{lh},k_\phi=0\right>$,$\left|\text{hh},k_\phi=1\right>$,$\left|\text{hh},k_\phi=2\right>$, and $\left|\text{hh},k_\phi=-2\right>$ in the basis. This way we have considered the exciton states, that are more likely to be excited with light and those states that primarily couple to them.

Results of our numerical calculation are discussed in the next section.

\section{Results}
\label{sec:results}

\subsection{Experimental results}

Figure~\ref{fig:NBE_overview} presents results of our experimental measurements. Absolute relative change of the nonradiative broadening is shown, with respect to the background nonradiative broadening value. Spectra have uniform vertical shift according to the magnetic field value, namely, the $\sigma^-$ spectra are shifted upwards, while the $\sigma^+$ ones are shifted downwards.  The light-hole exciton state dominates over the spectra on the left, showing small diamagnetic shift, which confirms significant Coulomb confinement of this state. Next to the right distinct states fan out with the magnetic field increase, while the steps in the spectra observed without the magnetic field gradually dissipate. This happens due to gradual degeneration of electron-hole states into the Landau levels. The lower distinguishable Landau level is well observed in magnetic fields in the 3-6\,T range (dashed lines). Apart from that, a particularly bright line can be distinguished, that is denoted by the solid line with arrows.

To reliably attribute the observed spectral features, we numerically calculated the energies of the Landau level states and of the exciton states under the magnetic field. The two approaches follow the same scheme described in subsection~\ref{sub:nummod}. Details for the Landau levels consideration are described in section~\ref{sub:app_Ll}. 

\subsection{Exciton states}

\begin{figure*}
    \centering
    \includegraphics[width=13.1cm]{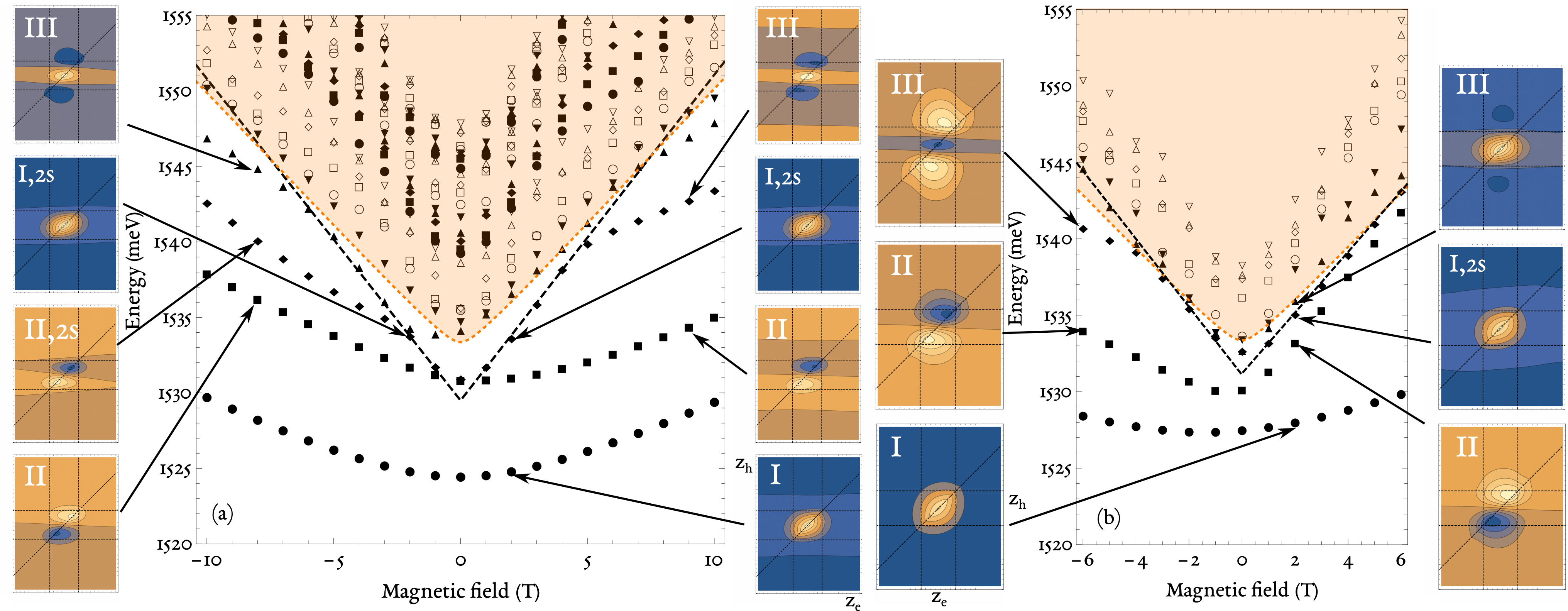}
    \caption{Calculated energies of the heavy-hole exciton states (a) and of the light-hole exciton states (b) in 14-nm GaAs/AlGaAs QW. Area shown in orange starts at the top of the QW barriers for holes. States in this area are likely to have a delocalized hole. Dashed V-shaped line is drawn along the excited exciton state obtained in the calculation. Colormaps illustrate exciton wave function amplitude distribution across $(z_e,z_h)$ plane at $\rho=0$. Dashed lines on colormaps denote the QW boundaries and $z_e=z_h$ diagonal.}
    \label{fig:exciton_calc}
\end{figure*}

Our second approach relies on the numerical solution of the Schroedinger equation for the exciton in the QW. We performed calculation for the 14-nm QW with 42-meV AlGaAs barriers. The valence band potential takes 33\% of the QW depth, leaving 67\% to the conduction band. Results for exciton calculation are shown in figure~\ref{fig:exciton_calc}. Panel (a) shows energies for the hh excitons, while panel (b) presents the lh ones. 

Analysing the obtained wave functions, we found that the states falling to the orange area are likely to have delocalized hole, which manifests itself in spreading of the wave function along the $z_h$ coordinate. The orange area starts at approximately $E=E_g+V_h$ at $B=0$\,T, where $E_g=1.519\,$eV is the GaAs band gap, and $V_h$ is the barriers height for the holes in our calculation. Generally speaking, the delocalized holes are still confined by Coloumb interaction with the electron in the QW, which should draw the energy of such states lower than the $E_g+V_h$. However, the  electrons are still a subject to the quantum-confined effect, which compensates for the Coulomb energy contribution to these states with the delocalized holes. Among the states with delocalized holes, there are many states with holes confined  by the numerical calculation domain along $z_h$ coordinate, which would not appear in the studied sample. Because of that we leave the orange region out of consideration and focus on the distinct states that are found below it.  

Analysing distinct exciton states below the orange area, we are able to distinguish the first, second and third states of the size quantization. In addition to that, the dashed line shows the state, which has a wave function slice at $\rho=0$ in $(z_e,z_h)$ plane similar to that of the ground state (see colormap "I,2s" in Fig.~\ref{fig:exciton_calc}(a)). The energy of this state exhibits the linear dependence on the magnetic field strength.

In our calculation, we spanned the magnetic field from $-10$ to $10\,$T for the hh excitons and from $-6$ to $6\,$T for the lh excitons. Since we composed the restricted basis so that the optically active exciton state has the magnetic momentum projection directed along the magnetic field axis, then the negative magnetic field is equivalent to the positive magnetic field for the exciton state with opposite magnetic momentum.

Maps around the fan diagram for exciton states in Fig.~\ref{fig:exciton_calc} depict a slices of the exciton wave functions at $\rho=0$ in coordinates $(z_e,z_h)$. The ground states for the hh and lh excitons are denoted as "I". Higher in energy, we found second state of size quantization, denoted as "II". We also found a hh exciton state "II, 2s", distinctively appearing below the orange area in the left-hand circular polarisation. Its wave function slice is similar to that of the "II" state. 

A third state of size quantization, denoted as "III", can be found at higher magnetic fields below the orange area. Finally there is a state with linear energy dependence on magnetic field, that we denoted as "I, 2s", obtained in both hh and lh exciton subsystems. We explore all distinct states in more detail in the next section.

\section{Discussion}

\begin{figure}
    \centering
    \includegraphics{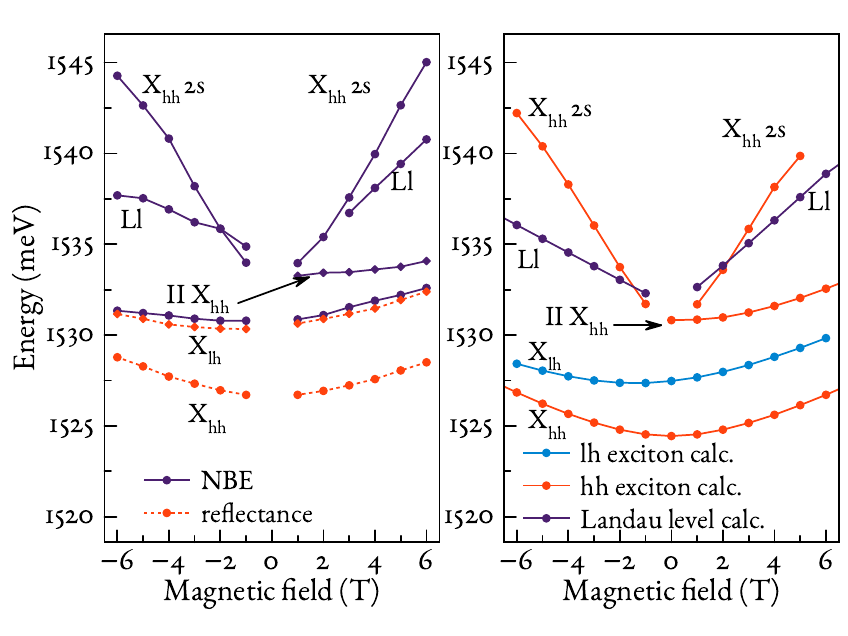}
    \caption{Comparison of the energies of main features extracted from the NBE and reflectance spectra (left), and obtained in the calculation (right) for the hh exciton subsystem, for the lh exciton subsystem, and for the free carriers subsystem. X$_\text{hh}$ and X$_\text{lh}$ denote the resonances related to the hh and lh exciton, respectively. "Ll" denote the lowest Landau level of the free electron-hole pair.}
    \label{fig:comparison}
\end{figure}

We compare our calculation results to the main features of the NBE and the reflectance spectra in figure~\ref{fig:comparison}. Similar to figure~\ref{fig:exciton_calc} the positive magnetic field corresponds to the coinciding direction of the field and the exciton angular momentum. So, switching circular polarisation is equivalent to switching of the magnetic field direction. The structure of the calculated states is similar to the experimentally observed structure. However, the energies at which the features appear are somewhat shifted. We assume that this shift comes from the deviation of the potential profiles in the studied sample from the nominal parameters used in the calculations. The calculations were performed for the 14-nm QW with 42 meV barriers. We believe that studied sample has slightly smaller QW width. This can be due to variation of Ga flow across the sample. A lower Ga flow can also explain the slightly increased barrier heights compared to our previous measurements of a different piece of the sample (see~Ref.\cite{Kurdyubov-PRB2021}). Indeed reduced Ga flow will lead to higher Al content in barriers. 

We would like to draw attention to the behavior of the second hh state of size quantization on the right-hand side in the figure~\ref{fig:exciton_calc}(a). It starts at vertex of the V-shaped dashed line and at $B=6$\,T draws closer to the ground lh exciton state shown in figure~\ref{fig:exciton_calc}(b). This region of the fan diagram has no other exciton states, which allow to ascribe the tiny peculiarity with similar behavior on experimental spectra to the second hh state of size quantization.

Typically, in narrow QW, the second state is considered undetectable, due to its even symmetry. Indeed, the exciton-light coupling constant (the radiative broadening), $\hbar\Gamma_0$, is governed by the absolute value of the overlapping integral:

\begin{equation}
    \hbar \Gamma_0 = \frac{1}{2}k_\text{ph} \omega_\text{LT} \pi a_B^3\left|\int \Phi(z) e^{i k_\text{ph} z}\, dz\right|^2.\label{gamma0}
\end{equation}
Here $k_\text{ph}$ is the light wave vector, $\omega_\text{LT}$ is the longitudinal transverse splitting, $a_B$ is the bulk exciton Bohr radius, and $\Phi(z)$ is the exciton wave function value along the line where electron and hole coordinate coincide. In slices presented in  figure~\ref{fig:exciton_calc}, this line is the dashed diagonal line.

The integral in expression~(\ref{gamma0}) has large absolute value for the even symmetry of  $\Phi(z)$ if $L_\text{QW}<<2\pi/k_\text{ph}$, and its value is small for the odd symmetry of the $\Phi(z)$ function. However it is not exactly zero. 

We estimated $\hbar \Gamma_0$ of the second hh exciton state of size quantization from our wave function calculation and found $\hbar \Gamma_0=0.6\,\mu$eV at $B=0$ and increases to $0.8\,\mu$eV at $B=6$\,T. The nonradiative broadening of the best samples we studied is comparable to the nonradiative broadening of the ground exciton state, which is about 40\,$\mu$eV for our sample. Therefore the reflectance spectra have no sign of resonances with radiative broadening value of about 1\,$\mu$eV. The NBE spectra, however, allows for observation of the second state of size quantization with $\hbar\Gamma_0$ much smaller than the typical nonradiative broadening.

\begin{figure*}
    \centering
    \includegraphics{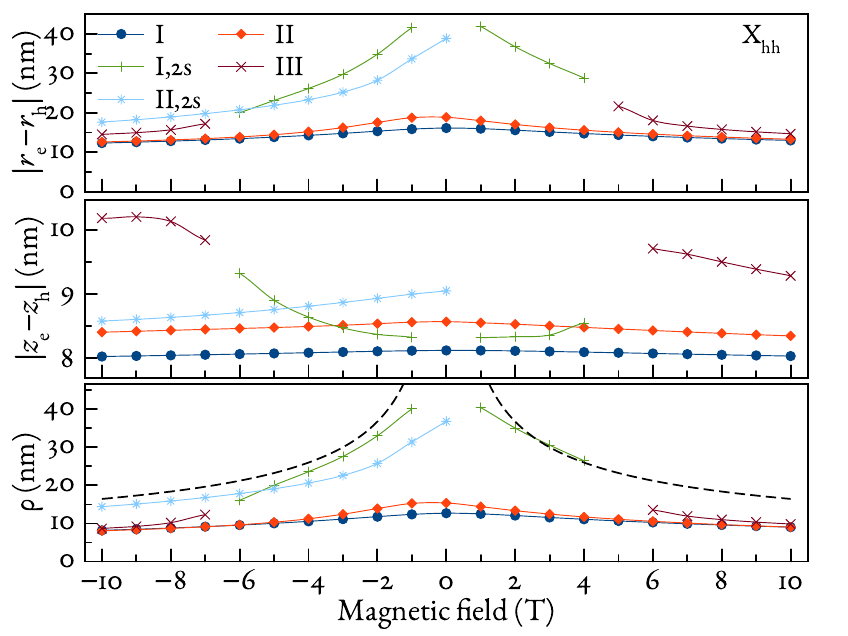}
    \includegraphics{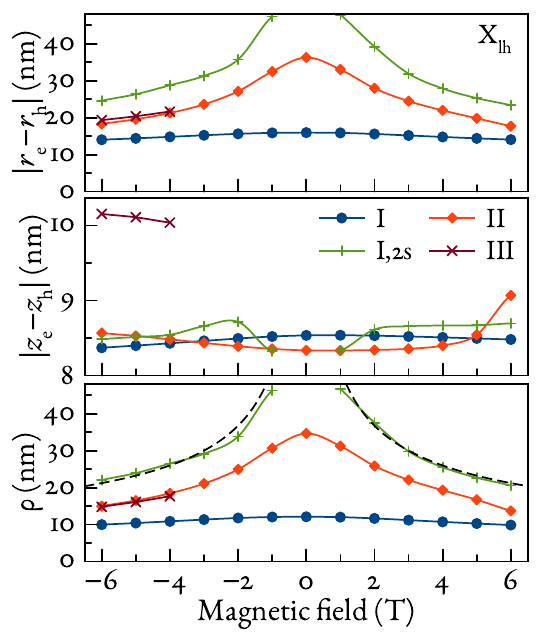}
    \caption{Mean electron-hole distances for the hh exciton states (left panel), and for the lh exciton states (right panel) obtained in the calculation. Panels from top to bottom demonstrate: the absolute relative electron-hole distance ($\left|r_e-r_h\right|$), the absolute electron-hole distance along $z$ direction ($\left|z_e-z_h\right|$), and perpendicular to that direction ($\rho$). Dashed line is the double magnetic length, $l_B=\sqrt{(c \hbar)/(e B)}$.}
    \label{fig:interp_hh} \label{fig:interp_lh}
\end{figure*}

To investigate further properties of the selected exciton states, we have calculated the electron-hole distance and its projections on $z$ and $\rho$ coordinates from the calculated wave functions. Figure~\ref{fig:interp_hh} present these values for distinct exciton states below the orange area in Fig.~\ref{fig:exciton_calc}. Mean electron-hole distance was calculated as an integral sum corresponding to the expression $\left<\psi\right|\left|r_e-r_h\right|\left|\psi\right>$, its projections were calculated likewise. Magnetic field confines the in-plane motion within the exciton states. The Coulomb coupling gets enhanced, which also leads to a decrease in the relative distance along direction $z$. However, the hh exciton 2s state has anticrossing with the third state of size quantization (III), resulting in the increase of the relative distance along $z$ direction for that state. The dashed line in the figures corresponds to the double magnetic length, 2$l_B$. It is remarkable how the relative in-plane distance. 

There are two factors, that play equally significant role in visibility of the exciton state in the NBE spectrum. They arise from the analysis of the exciton-light coupling constant expressed in equation~(\ref{gamma0}). (i) Function $\Phi(z)$ has larger amplitude for the exciton wave functions that posses smaller electron-hole distance, and, in addition, (ii) the symmetry of this function defines the overlapping with the light wave in the integral in Eq.~(\ref{gamma0}). For a QW with $L_\text{QW}<<2\pi/k_\text{ph}$, like in our case, even $\Phi(z)$ have significantly larger overlapping than an odd one.

From the top panel of figure~\ref{fig:interp_hh} we can distinguish more visible states from less visible based on the aforementioned factors. The lowest hh exciton state denoted as "I" is both has even $\Phi(z)$ and its electron-hole distance does not exceed 20\,nm. For the second state of size quantization ("II"), the electron-hole distance remains low, but $\Phi(z)$ is even. This state is barely visible in $\sigma^+$ polarisation, but it is untraceable in $\sigma^-$ compared to other "brighter" states appearing in it's vicinity. The function $\Phi(z)$ of the "I,2s" state is similar to that of the ground exciton state, but it has significantly larger electron-hole distance. This state is similar to the 2s Hydrogen state, and its larger electron-hole distance explains the diamagnetic shift, which increases faster than the Landau level of the unbound electron-hole transition. In the  calculation we also obtained "II,2s" state, which is visible below the orange region in $\sigma^-$ polarisation. It is both has odd symmetry of $\Phi(z)$ and larger electron-hole distance, therefore we could not observe this state in the experiments. We suppose that it is the 2s-like state for the second state of size quantization. Finally, there is a third state of size quantization, denoted as "III". In the calculations, it can be distinguished only at higher magnetic fields above $B=6\,$T, which lies outside of our experimental range. Experimentally we can not reliably trace this state as it falls in the orange area for all available magnetic fields. This state however experiences an anticrossing with the "I,2s" one. This is confirmed by the smooth transition of the electron-hole distances for these two states in Fig.~\ref{fig:interp_hh}.

The same logic can be applied to the lh exciton states considering a threefold lower matrix element of the dipole momentum for the lh subband. For the ground lh exciton state, denoted "I" in Figure~\ref{fig:interp_lh}, the electron-hole distance is barely larger than that of the ground hh exciton state, so we would expect roughly tree-times smaller exciton-light coupling constant. For the second size-quantized state (denoted "II"), the oscillator strength is suppressed by the symmetry of the $\Phi(z)$ function and furthermore three times reduced, therefore we expect it to be unobservable in the experiment. The "I,2s" state of the lh exciton appears to be close to that of the hh one with similar diamagnetic shift. But since it is also threefold weaker, we could not distinguish its input compared to the "I,2s" state of the hh exciton. Finally the third state of size quantization ("III") might be visible at $B=6\,$T in $\sigma^-$ polarisation, but its $\hbar \Gamma_0$ should be about 1\,$\mu$eV. Therefore it could produce only a small peculiarity slightly lower than the "I,2s" hh exciton state, which has significantly larger $\hbar \Gamma_0$.

\begin{figure}
    \centering
    \includegraphics{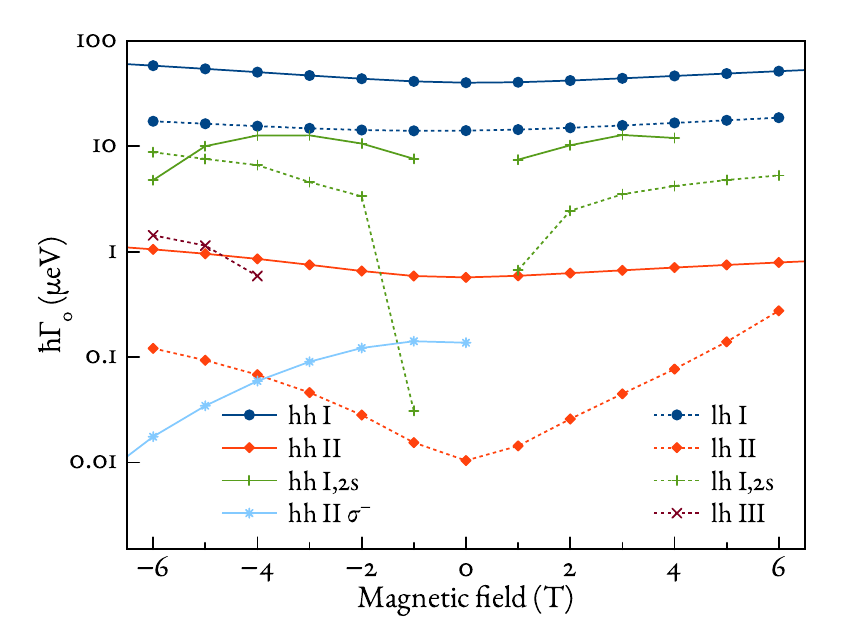}
    \caption{Oscillator strength for the distinct states obtained in the calculations. Solid/dashed lines correspond to hh/lh exciton states.  }
    \label{fig:calc_G0}
\end{figure}

Once we attributed all distinctive features of the observed NBE spectra, we could try to compare their radiative broadening constants obtained numerically to the feature amplitude in the NBE spectra. The numerically obtained $\hbar \Gamma_0$ are present in figure~\ref{fig:calc_G0}. As expected, "II,2s" hh state and "II" lh state posses the exciton-light coupling constants an order of magnitude smaller than those for the "II" hh state, which is barely observable in the NBE spectra. The ground lh state and "I,2s" hh state indeed should dominate in the NBE spectrum, when it excludes the ground hh state. Although the "I,2s" state of the lh exciton has comparatively small $\hbar\Gamma_0$ value, it increases with magnetic field, exceeding that of hh "I,2s" state. Given that these two states are close to each other, it is possible to misinterpret them in the NBE spectra. 

The quantitative analysis of our NBE spectra is hampered by the spectral resolution of our setup. Although we can accurately determine the central laser wavelength using wavelengthmeter, its resolution is limited by the linewidth of the tunable laser ($\Delta E=30\,\text{GHz}=0.13\,\text{meV}$). Indeed, the half-width at half-maximum (HWHM) of the features in our NBE spectra agrees well with this value. On the other hand the HWHM of the features should follow from the $\hbar\Gamma_0$ value of the exciton resonances, which would require very fine step laser tuning if its linewidth would be in MHz range, prolonging the measurement of each spectrum.

\section{Conclusion}

In conclusion, we have experimentally studied the nonradiative broadening excitation spectra of a GaAs/AlGaAs sample with a 14-nm QW under external magnetic field applied in the Faraday geometry. The spectra exhibited multiple resonances, which we identified through a microscopic modelling of energies and wave functions for various exciton transitions, and for transitions between Landau levels of free electrons and holes, taking into account the heavy-hole-light-hole coupling.

Experimental technique showed itself as extremely sensitive. We could follow the evolution of the second exciton state of size quantization, which has radiative decay rate constant as small as 1\,$\mu$eV. We could also observe a 2s-exciton state evolution. Experimental data showed good agreement with the numerical calculations. 

Moreover, we have calculated the energies of Landau levels with accounting of the heavy-hole-light-hole coupling and estimated the Coulomb correlation for the Landau levels with smallest energies and high angular momentum. We also observed transitions between these states in the experiment, with the theoretical treatment providing good agreement with the experimental results.

Finally, we found that the numerically obtained exciton-light coupling constants correlate with the signal amplitudes for the observed resonances, which confirms our interpretation of the experimental data. Overall, our study provides valuable insights into the electronic properties of GaAs-based nanostructures and their evolution under external magnetic field, with potential applications in the design and optimization of optoelectronic devices.

Developed numerical calculation method allows one to analyze inner structure of the observed exciton states such as electron-hole distance, spatial distribution of the wave function, and the exciton-light coupling strength.

\section*{Acknowledgement}

Authors are grateful to prof. I.V.\,Ignatiev for fruitful discussions. Financial support from the Russian Science Foundation, grant No. 21-72-00037, is acknowledged. The authors also thank Recourse Center “Nanophotonics” SPbU for the heterostructures studied in the present work and A. Levantovsky for the software “MagicPlot” extensively used for the data analysis.

\appendix

\section{NBE dependence on excitation power}
\label{sec:sample:appendix}

\begin{figure}
    \centering
    \includegraphics{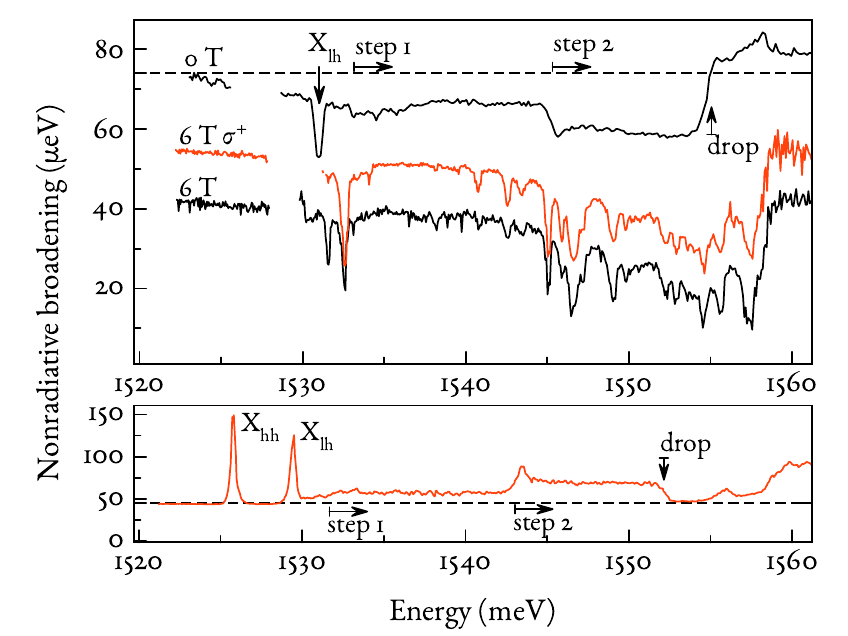}
    \caption{NBE spectra obtained in our measurement (top panel), and in earlier measurements described in the text (bottom panel). Features denoted as "step 1" and "step 2" correspond to the absorption of the additional excitation forming the free electron-hole pair.}
    \label{fig:NBE}
\end{figure}

Example of the NBE spectra obtained in our experiments are present in figure~\ref{fig:NBE}. The bottom panel presents the spectrum measured for the same heterostructure using a 80-femtosecond Ti:Sapphire laser for the reflectance spectrum detection and a CW Ti:Sapphire laser for the additional excitation. This measurement was made during our previous study~\cite{Kurdyubov-PRB2021}. We could direct the laser beams so that the stray light from the excitation beam did not affect the detection of the reflectance spectrum, and even the NBE signal from resonant excitation into the heavy-hole exciton could be collected.

In the current setup, however, a split coil magnet limits the angle aperture for light beams, with three cryostat windows contributing to the stray laser light, which results in a blind spot for the measurement of the NBE signal at the resonant excitation. The top panel of the figure~\ref{fig:NBE} presents the NBE spectra measured in magnetic cryostat at $B=0$\,T and at $B=6$\,T with selection of the right-circular polarisation (red curve) and without the selection. In this panel black curve  marked as "6 T" is shifted down to show the reproducibility of the spectrum. Although the curves "6 T" and "6 T $\sigma^+$" are a bit noisy, they demonstrate good reproducibility of spectral features present in the $\sigma^+$-polarisation. We can call this reproducibility to be exceptional, given that the curves were measured at different measurement days.

There is principal difference of the NBE spectra shown in the top and bottom panels of Fig.~\ref{fig:NBE}. The spectrum shown in the bottom panel exhibits increase in $\hbar\Gamma$ at each exciton resonance from relatively low background value of 47\,$\mu$eV. Above the resonances, two step-like increases of the broadening are observed, the first one at the energies $E>1530$\,meV, and the second one starting at 1543\,meV. We ascribe these two steps to the increase in the density of states of the free carriers. Once the excitation reaches the step, it excites free electron-hole pairs, which scatter the ground exciton increasing the nonradiative broadening. Finally, a drop appears at 1553\,meV, which corresponds to the energy when one of carriers is no longer confined by the QW potential. The "drop" feature spanning from 1553\,meV to 1560\,meV was explained in our previous work~\cite{Kurdyubov-PRB2021}. Starting from 1560\,meV, the absorption in the barrier layers contributes to the increase of the signal in the NBE spectrum.


In our current measurements (top panel) we use a different piece of the same heterostructure, that has a smaller QW width. This manifests itself in higher energies for the exciton states observed in the reflectance spectrum as well as in the NBE spectrum. A similar structure of the NBE spectrum is observed, but the nonradiative broadening in this case is decreasing at the light-hole exciton resonance and at each step feature. The "drop" is also observed as an inverted feature. This inversion, as we demonstrate below, comes from the difference in the reflectance measurement technique. Namely, the incandescent lamp used in these measurements noticeably contributes to the nonradiative broadening of exciton resonance (compare dashed line for the background broadening on the top panel of Fig.~\ref{fig:NBE} with that on the bottom one). 

\begin{figure}
    \centering
    \includegraphics{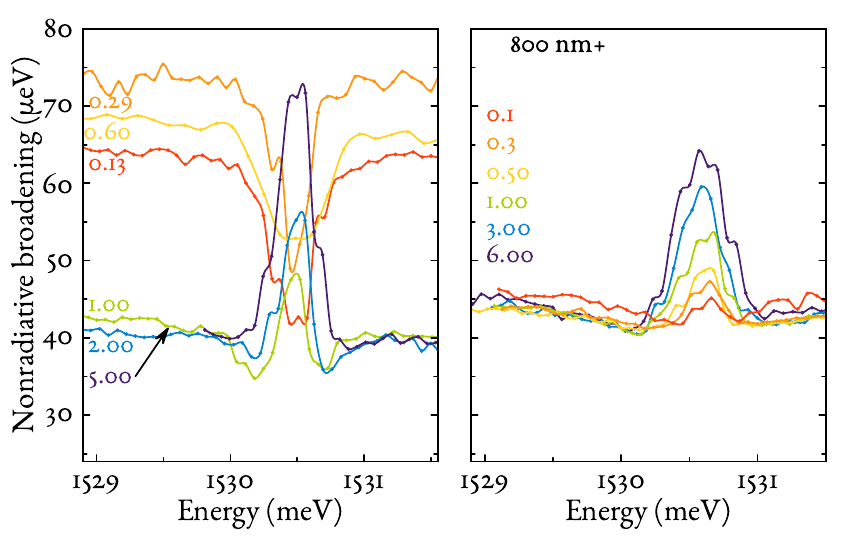}
    \caption{NBE signal at lh exciton excitation at various laser powers. Various colors correspond to laser powers from 0.1 to 6 mW. Data on the right panel were measured with the light of the light of the lamp filtered by the high-pass filter with edge at 800\,nm wavelength.
    \label{fig:app_power}}
\end{figure}

We have found that the excess broadening is caused by the short wavelength part of the incandesced lamp spectrum. Indeed, when we filtered the lamp light by a high-pass filter with the cut off edge at 800\,nm (1549\,meV), we obtained a "standard" NBE spectrum of the lh exciton similar to that observed earlier in Ref~\cite{Kurdyubov-PRB2021}. Figure~\ref{fig:app_power} demonstrates the transformation of the inverted NBE spectrum of the lh exciton resonance to the "normal" one at the increase of the CW laser power and at the usage of the cut off filter. 

We should note that the mechanism of this effect is still unclear. We can assume that the high-energy photons create free charge carriers in barrier layers and some fraction of these carriers are localized at deep centers. This causes a fluctuation electric field in the QW layer. The strong enough excitation by the CW laser creates some amount of free charged carriers in the QW layer, which screen the fluctuating electric field thus decreasing the broadening. In any case, this effect turned out to be useful for us because it allowed us to detect the NBE spectra using a semiconductor CW laser with small output power.

\section{Numerical method details}

Our numerical technique solves Schr\"{o}dinger equation for multidimensional Hamiltonian. A finite-difference method is used to convert differential operators into a sparse matrix. Finite-difference scheme accounts for the nonuniform grid, which is introduced to reduce number of grid nodes. The scheme was derived from the Tailor series expansion with technique, that was described in~\cite{Liu-AMM1995}. The matrix is either stored in Harwell-Boeing format for further use, or used within the mathematical package as is. Partial spectrum of the matrix is calculated using the Arnoldi algorithm. Obtained eigenvalues and eigenvectors we treat as energies and wave function approximations of the initial problem.

\subsection{Landau level numerical calculation}
\label{sub:app_Ll}
Three numerical problems were solved within the scope of this paper. First we calculated energies and wave functions of the Landau levels. The underlying logic here is based on the observation of step-like features in the NBE spectrum without magnetic field applied. Indeed, if absorption at the unbound electron-hole pairs is capable to change the nonradiative broadening of the ground hh exciton resonance, we assume that similarly, unbound carriers forming Landau levels would have absorption, performing similar effect.

A free electron in the bulk semiconductor subject to magnetic field has analytical solution for the Landau levels, which are harmonic oscillator mode with cyclotron frequency setting the scale of the spectrum, and infinitely degenerate states with two quantum numbers, conventionally called main quantum number and angular quantum number~\cite{Landau-3}.

\begin{figure}
    \centering
    \includegraphics[width=8.6cm]{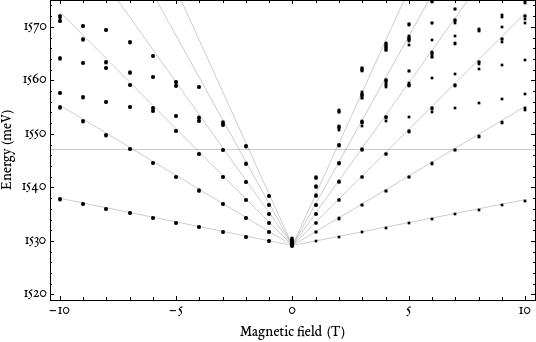}
    \caption{Energies of Landau levels calculated for 14-nm QW with 28 meV(2/3 of the 42\,meV full QW depth) barriers height. Horizontal line shows barriers band gap energy or electron. Fan diagram reproduces analytical solution according to formula~(\ref{Ll_electron})}.
    \label{fig:Ll_electron}
\end{figure}

With QW potential in place electron still has analytical solution, which however relies on the numerical procedure to sew wave function on the QW boundaries. For convinience we used our numerical method to obtain the electron Landau levels. The Hamiltonian for the electron in magnetic field in polar coordinates has the form:

\begin{eqnarray}
\hat{H}_e=-\frac{\hbar^2}{2 m_e}\frac{\partial^2}{\partial \rho^2}+\frac{\hbar^2}{2 m_e \rho}\frac{\partial}{\partial \rho}-\frac{\hbar^2}{2 m_e }\frac{\partial^2}{\partial z_e^2}\nonumber\\
+\left(\frac{e B}{2 c}\right)^2\frac{\rho^2}{2 m_e}-\frac{e B}{2 c}\frac{m \hbar}{m_e}+\frac{(m^2-1)\hbar^2}{2 m_e \rho^2}\nonumber\\
+U(z_e)+\frac{1}{2}g_e \mu_B B,
\end{eqnarray}
$m$ here is the angular quantum number. This expression is a result of substitution of generalized momentum~(\ref{mome}) into expression~(\ref{He}) and implementing ansatz for the electron wave function: $\phi(\rho,\phi,z_e)=\exp(i m \phi) \chi(z_e,\rho)/\rho$. We used division by $\rho$ to safely implement zero boundary conditions at $\rho=0$, taking into account that wave function is finite. As one can see the wave function is simply factorises in two parts, dependent on $z_e$ and $\rho$ coordinates separately. These parts correspond to states of size quantisation in the QW, and Landau level states. Energy of the Landau level for electron at the ground level of size quantisation has a form:

\begin{eqnarray}
    E_\text{Ll}=E_e+\left(n +\frac{1+m+|m|}{2}\right)\frac{\hbar e}{c m_e} B + \frac{1}{2} g_e \mu_B B,\label{Ll_electron}
\end{eqnarray}
where $m$ is a angular momentum of the landau level, and $n$ is a Landau level number. We calculated the wave functions using our numerical technique to benchmark our computation against the analytical solutions. We found that for the ground Landau level state difference between numerical and analitical solution does not exceed $\Delta E = 0.1$\,meV (see figure~\ref{fig:Ll_electron}). Calculation also captures additional series of Landau levels with energy above the barriers' band gap. Those are Landau levels for free electrons, which are subject to size quantisation effect within the calculation domain, which in our case if of about 50\,nm. Those states do not appear in our experiment, and we do not analyse them. 

The wave function analytically is constructed using the generalised Laguerre polynomials:

\begin{equation}
    \psi_e(\rho, \phi,z_e)=e^{i m \phi} \rho^{|m|} e^{-\frac{\rho^2}{4 l_B^2}} L_{-n}^{|m|+1}\left(\frac{\rho^2}{2 l_B^2}\right)\psi(z_e),
\end{equation}
where $\psi(z_e)$ is the electron wave function part defined by the square QW potential, and $l_B=\sqrt{ (c \hbar) / (e B)}$ is the magnetic length. One can see that Landau levels are infinitely degenerate, and states with larger numbers $m$ have larger distance to the coordinate origin.

For the holes however, Landau levels are subject to the hh-lh coupling, which we account using Luttinger Hamiltonian. The wave function for the holes when hh-lh coupling is ignored form same Landau levels in their part depending on the~$\rho$ coordinate, characterised by the same magnetic length, as only their energy depend on the effective mass of the particle. 

Let us consider electron and hole states with $n=0$ and $m$ corresponding to the ground Landau level energy. Due to the difference in the charge sign, in order to the charge carriers to be simultaneously on the ground Landau level energy, their numbers $m$ should have opposite signs. And for the transition between them to be possible the absolute values of $m$ should be the same. Therefore for the transition corresponding to the ground Landau levels for charge carriers the wave functions for electrons and holes have the form:

\begin{eqnarray}
    \psi_e(\rho_e, \phi_e,z_e)=\frac{e^{i m \phi_e} \rho_e^{|m|} e^{-\frac{\rho_e^2}{4 l_B^2}}}{\sqrt{2\pi 2^{|m|} |m|! l_B^{2(1+|m|)}}} \psi(z_e),\\
    \psi_h(\rho_h, \phi_h,z_h)=\frac{e^{-i m \phi_h} \rho_h^{|m|} e^{-\frac{\rho_h^2}{4 l_B^2}}}{\sqrt{2\pi 2^{|m|} |m|! l_B^{2(1+|m|)}}} \psi(z_h),
\end{eqnarray}

Since the probability density for electrons and holes are identical and not correlated with each other, for each $m$ the probability of absorption would remain the same. On the other hand, the Coulomb interaction matrix element in the form:

\begin{eqnarray}
    M_C=\int{-\frac{e^2}{\varepsilon \left|\mathbf{r}_e-\mathbf{r}_h\right|}\psi^*_e\psi^*_h\psi_e\psi_h\,d\mathbf{r}_ed\mathbf{r}_h,
    }
\end{eqnarray}
whould have different values for different $m$.

\begin{figure}
    \centering
    \includegraphics[width=8.6cm]{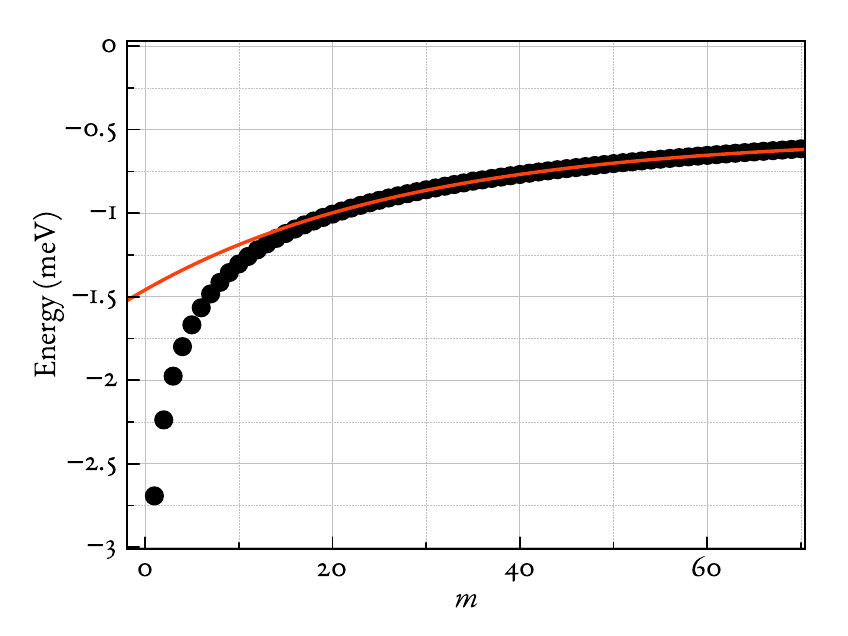}
    \caption{Coulomb interaction matrix element for the electron-hole pairs on the ground Landau levels with various values $m$ calculated at $B=1$\,T. Red line is a fitting data starting from $m=20$ with formula $M_C=a-c e^{-m/b}$. Best fit corresponds to $a=-0.53$\,meV, $b=28.8$, and $c=0.921$.}
    \label{fig:CoulombM}
\end{figure}

We calculated numerically this value for $B=1\,$T, and plotted it in Fig.~\ref{fig:CoulombM}. The matrix element saturates at value of about $-0.5$\,meV. Given that energy of the Landau level related to the magnetic field at magnetic field value $B=1$\,T is only~$E_\text{Ll}=1.52$\,meV we can not resort to perturbation theory. With magnetic field increase the $M_C$ growth inversely proportional to the magnetic length $l_B$, i.e. proportional to $\sqrt{B}$, while $E_\text{Ll}\propto B$. Therefore states with higher values of $m$ can become a valid starting point for a perturbation theory accounting for the Coulomb attraction as a small correction. From the numerical values present in figure~\ref{fig:CoulombM} we deduced a saturation value for the $M_C$ by fitting the data point using formula $M_C=a-c e^{-m/b}$, with $a=-0.53$\,meV. We use value $\delta E_C=-0.53 \sqrt{B}$\,meV with $B$ in Tesla to account for the Coulomb correlation of the Landau level energy. 

States on the Landau level with low value $m$ are affected by the Coulomb interaction beyond the perturbation approach. Indeed, these values contribute into forming of the exciton states, that are observed in our experiment and which we successfully describe using the numerical technique to calculate exciton states in magnetic field.

For the heavy-hole Landau level we also account for the hh-lh coupling. Similar to electron states we divide hole states in magnetic field according to their main quantum number $n$ and angular quantum number $m$. Similar to exciton, from this attribution we compose a selection rules for the hh-lh coupling of the hole states. The diagonal part of the Luttinger Hamiltonian has same structure as the one for the electron:

\begin{eqnarray}
\hat{H}_h=-\frac{\hbar^2}{2 m_{h\perp}}\frac{\partial^2}{\partial \rho^2}+\frac{\hbar^2}{2 m_{h\perp} \rho}\frac{\partial}{\partial \rho}-\frac{\hbar^2}{2 m_{h\parallel} }\frac{\partial^2}{\partial z_h^2}\nonumber\\
+\left(\frac{e B}{2 c}\right)^2\frac{\rho^2}{2 m_{h\perp}}+\frac{e B}{2 c}\frac{m \hbar}{m_{h\perp}}+\frac{(m^2-1)\hbar^2}{2 m_{h\perp} \rho^2}\nonumber\\
+U(z_h)+J_z 2 \kappa \mu_B B,\label{hh_diag}
\end{eqnarray}
here $J_z$ is the angular momentum projection of the hole, $J_z=\pm3/2$ for the heavy hole, and $J_z=\pm1/2$ for the light hole. Effective masses are related to the Luttinger parameters as follows:

\begin{eqnarray}
    m_{\text{hh}\parallel}=\frac{m_0}{\gamma_1-2\gamma_2}\quad m_{\text{hh}\perp}=\frac{m_0}{\gamma_1+\gamma_2}\nonumber\\
    m_{\text{lh}\parallel}=\frac{m_0}{\gamma_1+2\gamma_2}\quad m_{\text{lh}\perp}=\frac{m_0}{\gamma_1-\gamma_2}\nonumber
\end{eqnarray}

Similar anzats that we used for electron states produces an infinite Hamiltonian matrix with operator~(\ref{hh_diag}) on the diagonal. We restrict hamiltonian matrix to a four Landau level states that primarily couple to the hh state with angular momentum $J_z=3/2$ and form a matrix:

\begin{equation}
\hat{H}^m=
    \begin{pmatrix}
        \hat{H}^m_{hh} & \hat{H}^{m+1}_{12} & \hat{H}^{m+2}_{13} & \hat{H}^{m-2}_{13} \\
        \hat{H}^{m-1}_{21} & \hat{H}^{m+1}_{lh} & 0 & 0 \\
        \hat{H}^{m-2}_{31} & 0 & \hat{H}^{m+2}_{lh} & 0 \\
        \hat{H}^{m+2}_{31} & 0 & 0 & \hat{H}^{m-2}_{lh}
    \end{pmatrix}
\end{equation}

\begin{figure}
    \centering
    \includegraphics{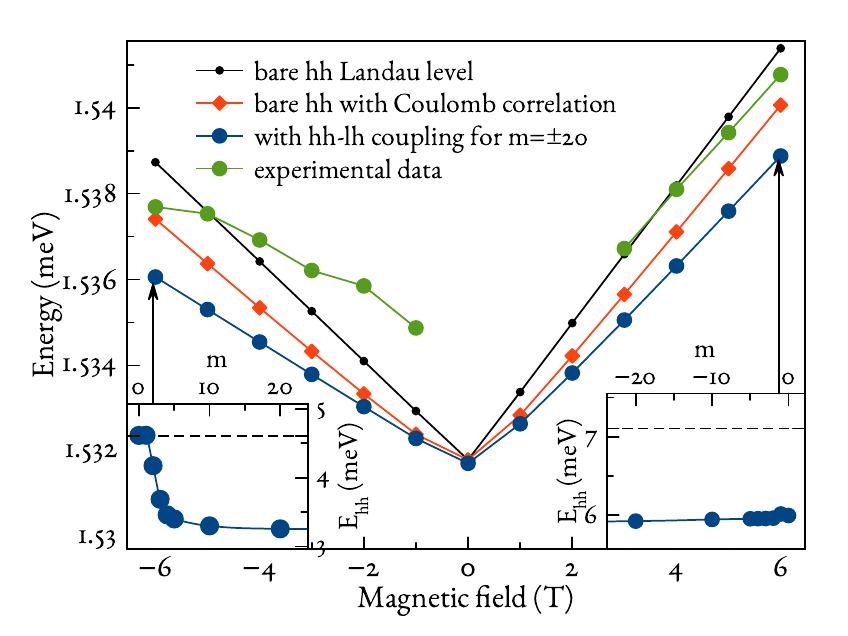}
    \caption{Energy of the ground Landau level at various magnetic fields. Energies for transitions between electron with spin projection $\sigma=-1/2$ with bare hh (black), bare hh with Coulomb correlation accounted (red), and hh state with hh-lh coupling accounted for $m=\pm20$ in addition to Coulomb correlation (blue). hh states have angular momentum projection $J_z=3/2$. Green points are the ground Landau level energies extracted from experiment. Insets show the hh energy shift induced by the hh-lh coupling for states with various $m$ at $B=6\,$T. Grid lines in the insets correspond to the bare hh energy, energies are shown relative to the QW bottom.}
    \label{fig:app_Ll_compare}
\end{figure}

The non-diagonal components of the Hamiltonian, taking into account chosen coordinate system, symmetric gauge and ansatz for the wave function, have the following form:

{\small
\begin{eqnarray}
    &H^{k+1}_{12}=\frac{\sqrt{3}\hbar^2\gamma_3}{m_0}i \left[\left(\frac{e B}{2 c \hbar}\rho+\frac{(k-1)}{\rho}\right)\frac{\partial}{\partial z_h}+\frac{\partial}{\partial \rho}\frac{\partial}{\partial z_h}\right]& \label{nondiag_Ll12}\\
    &H^{k-1}_{21}=\frac{\sqrt{3}\hbar^2\gamma_3}{m_0}i \left[\left(\frac{e B}{2 c \hbar}\rho+\frac{(k+1)}{\rho}\right)\frac{\partial}{\partial z_h}-\frac{\partial}{\partial \rho}\frac{\partial}{\partial z_h}\right] \label{nondiag_Ll21}&
\end{eqnarray}
\begin{eqnarray}
    &\hat{H}^{k+2}_{13}=\frac{\sqrt{3}(\gamma_2-\gamma_3)}{2}\left(\frac{ \hbar ^2}{2 m_0}\frac{\partial^2}{\partial \rho^2}
    +\frac{ \hbar \left(B e \rho ^2-3 c \hbar +2 c k \hbar \right)}{2 c m_0 \rho }\frac{\partial}{\partial \rho}\right.\nonumber\\
    &\left.+\frac{ \left(B^2 e^2 \rho ^4+4 B c e \rho ^2 \hbar  (-1+k)+4 c^2 \hbar ^2 \left(\left(k^2+3\right)-4 k\right)\right)}{8 c^2 m_0 \rho ^2}\right) \label{nondiag_Ll13k2}
\end{eqnarray}
\begin{eqnarray}
    &\hat{H}^{k-2}_{13}=
    \frac{\sqrt{3}(\gamma_2+\gamma_3)}{2}\left(\frac{ \hbar ^2}{2 m_0}\frac{\partial^2}{\partial \rho^2}
    -\frac{ \hbar \left(B e \rho ^2+3 c \hbar +2 c k \hbar \right)}{2 c m_0 \rho }\frac{\partial}{\partial \rho}\right.\nonumber\\
    &\left.+\frac{ \left(B^2 e^2 \rho ^4+4 B c e \rho ^2 \hbar  (1+k)+4 c^2 \hbar ^2 \left(\left(k^2+3\right)+4 k\right)\right)}{8 c^2 m_0 \rho ^2}\right)\label{nondiag_Ll13km2}
\end{eqnarray}
\begin{eqnarray}
    &\hat{H}^{k+2}_{31}=\frac{\sqrt{3}(\gamma_2+\gamma_3)}{2}\left(\frac{ \hbar ^2}{2 m_0}\frac{\partial^2}{\partial \rho^2}
    +\frac{ \hbar \left(B e \rho ^2-3 c \hbar +2 c k \hbar \right)}{2 c m_0 \rho }\frac{\partial}{\partial \rho}\right.\nonumber\\
    &+\left.\frac{ \left(B^2 e^2 \rho ^4+4 B c e \rho ^2 \hbar  (-1+k)+4 c^2 \hbar ^2 \left(\left(k^2+3\right)-4 k\right)\right)}{8 c^2 m_0 \rho ^2}\right)\label{nondiag_Ll31k2}
\end{eqnarray}
\begin{eqnarray}
    &\hat{H}^{k-2}_{31}=\frac{\sqrt{3}(\gamma_2-\gamma_3)}{2}\left(\frac{ \hbar ^2}{2 m_0}\frac{\partial^2}{\partial \rho^2}
    -\frac{ \hbar \left(-B e \rho ^2+3 c \hbar -2 c k \hbar \right)}{2 c m_0 \rho }\frac{\partial}{\partial \rho}\right.\nonumber\\
    &\left.+\frac{ \left(B^2 e^2 \rho ^4+4 B c e \rho ^2 \hbar  (1+k)+4 c^2 \hbar ^2 \left(\left(k^2+3\right)+4 k\right)\right)}{8 c^2 m_0 \rho ^2}\right)\label{nondiag_Ll31km2}
\end{eqnarray}
  }
With constructed restricted Hamiltonian matrix we composed a sparse matrix using finite-difference method on the $60\times60$ grid in the $65\,\text{nm}\times400\,\text{nm}$ domain with nonuniform grid. We calculated eigenvalues of the matrices for various magnetic field spanning from $B=-10\,$T to $B=10\,$T with 1\,T step using Arnoldi algorithm. We performed this calculation for various $m$ of the hh state. For the electron states with various angular momenta on the lowest Landau level have identical energies, but for the hole these energies slightly differ due to varying hh-lh coupling. Figure~\ref{fig:app_Ll_compare} summarises our investigation of the ground Landau level. We compose
Landau level energy as a sum of the GaAs band gap $E_g=1519$\,meV, calculated electron Landau level energy with spin projection~$\sigma=-1/2$, calculated energy of the hh state with magnetic momentum progaction~$J_z=3/2$ and $m=\pm20$(the sign depends on the magnetic field direction), and Coulomb correlation energy previously defined as $M_C(B)=-0.53\sqrt{B}$\,meV, with $B$ in T. Opposite direction of electron and hole magnetic momenta is necessary to fulfill the selection rules present in figure~\ref{fig:scheme}. As seen from the insets in figure~\ref{fig:app_Ll_compare}, the hh-lh coupling driven energy shift saturates at large absolute values of $m$ similar to the saturation of the Coulomb matrix element~(see Fig.\ref{fig:CoulombM}). We therefore come to the conclusion that all states with large $m$ compose a single absorption level, which energy should correspond to the blue line in figure~\ref{fig:app_Ll_compare}. Comparing to the experimentally observed energy values, we conclude that our calculated energies closely follow the incline of the experimental data, and are slightly shifted by the constant value. We explain this shift with slightly smaller QW width in the sample relative to the nominal parameters at which the calculation was carried out.

\subsection{Exciton states numerical calculation}

A similar technique was used to calculate the exciton enegies and wave functions. The Hamiltonian for the exciton combines the two Hamiltonians for the electron and for the hole, it accounts for the Coulomb coupling and for the hh-lh coupling. We used polar coordinates for the relative electron-hole distance in the QW plane and individual coordinates in the growth direction. The coordinate transform was $(x_e,y_e,z_e,x_h,y_h,z_h)\longleftrightarrow(X,Y,\rho,\phi,z_e,z_h)$. The resulting formulas are too large to fit into the paper, and therefore we provide them in supplementary file in the Wolfram Mathematica notebook format.

Used ansatz for the wave function, proposed by Gorjkov and Dzjaloshinsky~\cite{Gorjkov-ZETP1967} preserves the form of the operators in the Luttinger Hamiltonian. We constructed a restricted Hamiltonian similarly to the one used for the Landau level calculation:

\begin{widetext}
\begin{equation}
\hat{H}_{ex}=
    \begin{pmatrix}
        \hat{H}_{hh, k_\phi=0} & \hat{H}_{12} & \hat{H}_{13, \delta k_\phi=2} & \hat{H}_{13,\delta k_\phi=-2} \\
        \hat{H}_{21} & \hat{H}_{lh, J_z=1/2, k_\phi=1} & 0 & 0 \\
        \hat{H}_{31,\delta k_\phi=2} & 0 & \hat{H}_{lh, J_z=-1/2, k_\phi=2} & 0 \\
        \hat{H}_{31,\delta k_\phi=-2} & 0 & 0 & \hat{H}_{lh, J_z=-1/2, k_\phi=-2}
    \end{pmatrix}
\end{equation}
\end{widetext}

The diagonal part of this Hamiltonian with chosen ansatz applied has the form:

\begin{eqnarray}
    &\hat{H}_{hh}=-\frac{\hbar^2}{2 m_{\parallel hh}}\frac{\partial^2}{\partial z_h^2}-\frac{\hbar^2}{2 m_e}\frac{\partial^2}{\partial z_e^2}+U(z_e,z_h)&\nonumber\\
    &-\frac{\hbar^2}{2 \mu_{hh}}\frac{\partial^2}{\partial \rho^2}+\frac{\hbar^2}{2 \mu_{hh}}\frac{\partial}{\partial \rho}+\frac{(k_\phi^2-1)\hbar^2}{2 \rho^2 \mu_{hh}}&\nonumber\\
    &-\frac{e B}{2c}\frac{\hbar (m_e-m_{hh})}{m_e m_{hh}}+\left(\frac{e B}{2 c}\right)^2\frac{\rho^2}{2 \mu_{hh}}&\nonumber\\
    &-\frac{1}{2}g_e \mu_B B - 2\kappa\frac{3}{2} \mu_B B,&    
\end{eqnarray}

\begin{eqnarray}
    &\hat{H}_{lh}=-\frac{\hbar^2}{2 m_{\parallel lh}}\frac{\partial^2}{\partial z_h^2}-\frac{\hbar^2}{2 m_e}\frac{\partial^2}{\partial z_e^2}+U(z_e,z_h)&\nonumber\\
    &-\frac{\hbar^2}{2 \mu_{lh}}\frac{\partial^2}{\partial \rho^2}+\frac{\hbar^2}{2 \mu_{lh}}\frac{\partial}{\partial \rho}+\frac{(k_\phi^2-1)\hbar^2}{2 \rho^2 \mu_{lh}}&\nonumber\\
    &-\frac{e B}{2c}\frac{\hbar (m_e-m_{lh})}{m_e m_{hh}}+\left(\frac{e B}{2 c}\right)^2\frac{\rho^2}{2 \mu_{lh}}&\nonumber\\
    &-\frac{1}{2}g_e \mu_B B + 2\kappa J_z \mu_B B.&    
\end{eqnarray}
Here $\mu_{hh,lh}$ is the reduced mass with $m_{\perp hh,lh}$ hole mass. The nondiagonal parts of the restricted Hamiltonian have the form:

{\small
\begin{eqnarray}
    &H_{12}=\frac{\sqrt{3}\hbar^2\gamma_3}{m_0}i \left[\left(\frac{e B}{2 c \hbar}\rho+\frac{(1-k_\phi)}{\rho}\right)\frac{\partial}{\partial z_h}-\frac{\partial}{\partial \rho}\frac{\partial}{\partial z_h}\right]&\\
    &H_{21}=\frac{\sqrt{3}\hbar^2\gamma_3}{m_0}i \left[\left(\frac{e B}{2 c \hbar}\rho-\frac{(1+k_\phi)}{\rho}\right)\frac{\partial}{\partial z_h}+\frac{\partial}{\partial \rho}\frac{\partial}{\partial z_h}\right]&
\end{eqnarray}
\begin{eqnarray}
    &\hat{H}_{13, \delta k_\phi = 2}=\frac{\sqrt{3}(\gamma_2-\gamma_3)}{2}\left(\frac{ \hbar ^2}{2 m_0}\frac{\partial^2}{\partial \rho^2}
    -\frac{ \hbar \left(B e \rho ^2+3 c \hbar -2 c k \hbar \right)}{2 c m_0 \rho }\frac{\partial}{\partial \rho}\right.\nonumber\\
    &\left.+\frac{ \left(B^2 e^2 \rho ^4-4 B c e \rho ^2 \hbar  (-1+k_\phi)+4 c^2 \hbar ^2 \left(\left(k_\phi^2+3\right)-4 k_\phi\right)\right)}{8 c^2 m_0 \rho ^2}\right)
\end{eqnarray}
\begin{eqnarray}
    &\hat{H}_{13, \delta k_\phi = -2}=
    \frac{\sqrt{3}(\gamma_2+\gamma_3)}{2}\left(\frac{ \hbar ^2}{2 m_0}\frac{\partial^2}{\partial \rho^2}
    +\frac{ \hbar \left(B e \rho ^2-3 c \hbar -2 c k_\phi \hbar \right)}{2 c m_0 \rho }\frac{\partial}{\partial \rho}\right.\nonumber\\
    &\left.+\frac{ \left(B^2 e^2 \rho ^4-4 B c e \rho ^2 \hbar  (1+k_\phi)+4 c^2 \hbar ^2 \left(\left(k_\phi^2+3\right)+4 k_\phi\right)\right)}{8 c^2 m_0 \rho ^2}\right)
\end{eqnarray}
\begin{eqnarray}
    &\hat{H}_{31, \delta k_\phi = 2}=\frac{\sqrt{3}(\gamma_2+\gamma_3)}{2}\left(\frac{ \hbar ^2}{2 m_0}\frac{\partial^2}{\partial \rho^2}
    -\frac{ \hbar \left(B e \rho ^2+3 c \hbar -2 c k_\phi \hbar \right)}{2 c m_0 \rho }\frac{\partial}{\partial \rho}\right.\nonumber\\
    &+\left.\frac{ \left(B^2 e^2 \rho ^4-4 B c e \rho ^2 \hbar  (-1+k_\phi)+4 c^2 \hbar ^2 \left(\left(k_\phi^2+3\right)-4 k_\phi\right)\right)}{8 c^2 m_0 \rho ^2}\right)
\end{eqnarray}
\begin{eqnarray}
    &\hat{H}_{31, \delta k_\phi = -2}=\frac{\sqrt{3}(\gamma_2-\gamma_3)}{2}\left(\frac{ \hbar ^2}{2 m_0}\frac{\partial^2}{\partial \rho^2}
    +\frac{ \hbar \left(B e \rho ^2-3 c \hbar -2 c k \hbar \right)}{2 c m_0 \rho }\frac{\partial}{\partial \rho}\right.\nonumber\\
    &\left.+\frac{ \left(B^2 e^2 \rho ^4-4 B c e \rho ^2 \hbar  (1+k_\phi)+4 c^2 \hbar ^2 \left(\left(k_\phi^2+3\right)+4 k_\phi\right)\right)}{8 c^2 m_0 \rho ^2}\right)
\end{eqnarray}
  }
These expressions repeat operators for the free holes up to some signs change. This is due to relative coordinate definition, which differs from the polar coordinates for the hole in expressions~(\ref{nondiag_Ll12}--\ref{nondiag_Ll31km2}). In these expressions $\gamma_1=6.98$, $\gamma_2=2.06$, $\gamma_3=2.9$, and $\kappa=-1.2$ are the Luttinger parameters chosen according to the Ref.~\cite{Vurgaftman-JAP2001}. Electron $g$-factor is $g_e=-0.44$ in agreement with Roth's formula~\cite{Roth-PR1959, Yugova-PRB2007}. Masses~$m_{\parallel hh,lh}$ and $m_{\perp hh,lh}$ are defined by the Luttinger parameters according to expressions in the previous subsection.

 \bibliographystyle{elsarticle-num} 





\end{document}